\documentclass{IEEEtran4PSCC}
\ifCLASSINFOpdf
   \usepackage[pdftex]{graphicx}
\else
   \usepackage[dvips]{graphicx}
\fi

\usepackage[cmex10]{amsmath}

\usepackage{amssymb}
\usepackage{here}
\graphicspath{{./Figures/}}
\usepackage{color}
\usepackage[utf8]{inputenc}
\usepackage[T1]{fontenc}
\usepackage{enumerate}
\usepackage{booktabs}
\usepackage{epstopdf}
\usepackage{cite}
\usepackage{soul}
\usepackage{subfig}
\usepackage{mathtools}

\usepackage{algorithm}
\usepackage{algpseudocode}

\usepackage{tabularray}

\makeatletter
\let\old@ps@headings\ps@headings
\let\old@ps@IEEEtitlepagestyle\ps@IEEEtitlepagestyle
\def\psccfooter#1{%
    \def\ps@headings{%
        \old@ps@headings%
        \def\@oddfoot{\strut\hfill#1\hfill\strut}%
        \def\@evenfoot{\strut\hfill#1\hfill\strut}%
    }%
    \def\ps@IEEEtitlepagestyle{%
        \old@ps@IEEEtitlepagestyle%
        \def\@oddfoot{\strut\hfill#1\hfill\strut}%
        \def\@evenfoot{\strut\hfill#1\hfill\strut}%
    }%
    \ps@headings%
}
\makeatother

\begin{document}
\title{Computationally Efficient Electromagnetic Transient Power System Studies using Bayesian Optimization}

\author{
\IEEEauthorblockN{Marius Kuhn}
\IEEEauthorblockA{IAEW, RWTH Aachen \\
Germany\\
\ m.kuhn@iaew.rwth-aachen.de}
\and
\IEEEauthorblockN{Evelyn Heylen}
\IEEEauthorblockA{Mindbrew BV\\
Belgium\\
evelyn@evelynheylen.eu}
\and
\IEEEauthorblockN{Willem Leterme}
\IEEEauthorblockA{IAEW, RWTH Aachen \\
Germany\\
\ w.leterme@iaew.rwth-aachen.de}}

\maketitle

\begin{abstract}
The power system of the future will be governed by complex interactions and non-linear phenomena at small time-scales, that should be studied more and more through computationally expensive software simulations. To solve the abovementioned problems, power system engineers face problems with following characteristics: (i) a computationally expensive simulator, (ii) non-linear functions to optimize and (iii) lack of abundance of data. Existing optimization settings involving EMT-type simulations have been developed, but mainly use a deterministic model and optimizer, which may be computationally inefficient and do not guarantee finding a global optimum. Furthermore, the main focus has been on optimization routines, and less attention has been paid to other tasks such as classification. In this paper, an automation framework based on Bayesian Optimization is introduced, and applied to two case studies involving optimization and classification. It is found that the framework has the potential to reduce computational effort, outperform deterministic optimizers and is applicable to a multitude of problems. Nevertheless, it was found that the output of the Bayesian Optimization depends on the number of samples used for initialization, and in addition, careful selection of surrogate models, which should be subject to future investigation.
\end{abstract}

\begin{IEEEkeywords}
Bayeisan Optimization, Electromagnetic Transients, Classification
\end{IEEEkeywords}


\section{Introduction}
\label{sec:introduction}

The power system of the future will be characterized by complex interactions that should be studied more and more through computationally expensive software simulations. The increased complexity may be in the form of control interactions and stability~\cite{gu_power_2023}, increased occurrence of resonance phenomena in undergrounded cable systems~\cite{cigre_wg_c4307_transformer_2014}, increasingly complex protection function analysis or complex phenomena in HVDC grids, e.g., affecting the design of HVDC circuit breakers~\cite{jahn_holistic_2021}. To study these problems, often involving non-linearities, extensive models are being developed to run in EMT-type or real-time simulators, leading to highly computationally expensive simulations. It will be tedious to run all these simulations, given their computational cost together with the sheer number of possible power system states.

In the recent past, several fields have started using elements from artificial intelligence and machine learning to replace human efforts. Notable examples are the Robot Scientist~\cite{king_robot_2005}, use of machine learning in physics~\cite{baydin_efficient_2019}, or active discovery of organic semiconductors~\cite{kunkel_active_2021}. In all cases, machine learning is used to replace human effort in devising a sequence of simulations considering maximum outcome with minimal cost of  expensive hardware experiments or computationally expensive software simulations. In essence, all these cases are similar in following components: a non-linear function to be optimized (i.e., solved by the expensive simulator), a surrogate model, and a function that sequentially selects the next simulation by acting on the surrogate model.

Many power system problems can be cast into the setting as described above. The expensive simulator consists in this case of an extensive dynamic or electromagnetic transient (EMT) model together with an EMT-type solver. Non-linear functions to optimize are, e.g., finding the maximum overvoltage that occurs during switching, finding the probability of exceeding a threshold during switching, finding an optimal control setting, finding an optimal equipment size. Several of these settings may even involve probabilistic analysis, which could yield a large number of simulations per function evaluation.


Past research has focused on combining optimization methods with EMT-type simulators, where the effectiveness of the approach depends on the metaheuristic that is being used. In~\cite{gole_optimization-enabled_2005} and~\cite{fratila_stability_2019}, single-solution approaches have been used, which have the advantage of a lower number of required simulations at the cost of potentially not reaching the global optimum. In~\cite{jahn_holistic_2021} and~\cite{nzale_tool_2023}, population-based methods have been used that have a higher probability of achieving the global optimum while also requiring a larger number of simulations (e.g., in the range of twenty per optimization step). In~\cite{tanaka_application_2018}, an optimization method is devised by combining Kriging models with the Expected Improvement-heuristic. This is a special case of Bayeisan Optimization, upon which we extend in this paper.

In this paper, we show how the framework of Bayesian Optimization (BO) can be applied to efficiently automate complex power system studies. The paper extends on the state-of-the-art, by covering the application of BO in detail and by applying it beyond optimization tasks. The contribution of this paper is thus three-fold: (i) an introduction to the application of BO to different types of problems, (ii) comparison of different implementations and uses of BO and (iii) a comparison of BO to other optimization methods in the context of power system simulations.

The paper first describes the settings in which automation can take place, before discussing the elements of the automation framework. Then, the paper discusses the implementation of the BO framework in Section IV, prior to applying it in Section V. In Section V, the BO framework is additionally compared against other algorithms. Conclusions are drawn in Section VI.

\section{Automation Settings}

\begin{figure*}[t]
  \centering
  \subfloat[BO]{\includegraphics{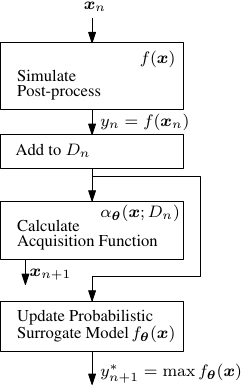}}~~~~
  \subfloat[Active Learning]{\includegraphics{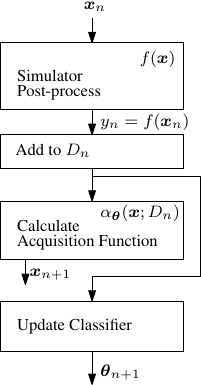}}~~~~
  \subfloat[Active Search]{\includegraphics{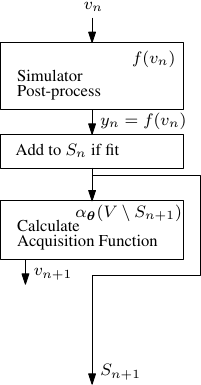}}~~~~
  \subfloat[Active Discovery]{\includegraphics{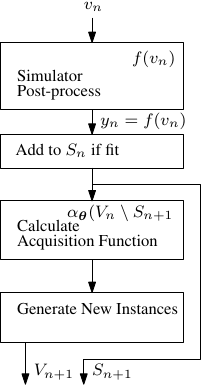}}
  \caption{Workflow for Bayesian Optimization (a), Active Learning (b), Active Search (c) and Active Discovery (d).}
  \label{fig:workflows}
\end{figure*}

\subsection{Optimization}
\label{sec:optimization}

In the field of optimization, BO is a global optimization method suitable for problems where an optimal value $y^*$ must be found for a function $y=f(\boldsymbol{x})$ that is expensive to evaluate or may lack an exact functional form (Fig.~\ref{fig:workflows}a). The optimization requires a probabilistic surrogate model $f_{\boldsymbol{\theta}}(\boldsymbol{x})$ and an acquisition function $\alpha_{\boldsymbol{\theta}}(\boldsymbol{x};D_n)$ that acts on the previously evaluated points, collected in $D_n$. The acquisition function steers function evaluations to those with the highest reward (in this sense, revealing most information related to the optimum). As such, in~\cite{ghahramani_probabilistic_2015}, BO is linked to reinforcement learning in systems where decisions do not affect the outcomes. The outcome of the optimization is $y^* = \max_{\boldsymbol{x}} f_{\boldsymbol{\theta}}(\boldsymbol{x})$.


In power systems, many problems can be cast into an optimization problem with unknown functional form, especially those involving time-domain simulations. Notable examples that have been tackled in the literature are: design of breakers and inductors in HVDC grids~\cite{jahn_holistic_2021}, design of controls for stability, matching of simulator parameters to real-life measurements~\cite{fratila_stability_2019}, or optimizing converter controls for power system protections \cite{leterme_rethinking_2023}. These examples typically include a highly expensive function to simulate and the absence of functional forms, whereas the number of dimensions is relatively limited.  

\subsection{Classification}
\label{sec:classification}


In classification settings, active learning is used to find the parameters $\boldsymbol{\theta}$ of a generalizable classifier while minimizing the number of labeling actions in a set of unlabeled instances. In the traditional setting, the input to the problem are unlabeled instances that have been collected, while the expensive function $y=f(\boldsymbol{x})$ represents the human expert or machine that has to label the instances~\cite{settles_active_2012}. Here, a probabilistic classifier $f_{\boldsymbol{\theta}}(\boldsymbol{x})$ together with an acquisition function $\alpha_{\boldsymbol{\theta}}(\boldsymbol{x};D_n)$ are used, while the aim is to maximize classifier generalizability. Active learning has been used in conjunction with probabilistic classifiers but also with support vector machines~\cite{kremer_active_2014}.


Applications of active learning in power systems may play a role in the field of power system protection, power quality monitoring or asset management. With an increasing amount of data collected from sensors in the system, automated tools will be required for analysis, e.g., to indicate when human expert input is required. Furthermore, active learning may be used in cases where existing and known system states need to be classified. The active learning procedure will aid in selecting only those cases that require human expertise or expensive simulations. The case studied in this paper uses active learning to classify instances with respect to a threshold that results in equipment failure.

\subsection{Active Search and Discovery}
\label{sec:search}

In active search or discovery settings, the aim is to maximize the number of instances found or created that fulfill certain criteria or fitness functions~\cite{ma_active_2015}. These methods may encode information on instances using graphs, representing instances as nodes $v_n \in V$, and similarities between instances using edges between nodes. The quality of the instances is described by a fitness function $y=f(v_n)$. A probabilistic model together with an acquisition function $\alpha_{\boldsymbol{\theta}}(V \setminus S_n)$ directs the search in the graph, or the direction in which new instances should be created. It eliminates the visited edges $S_n$ from the search, to find as many instances as possible that fulfill the fitness function (active search) or create new meaningful instances (active discovery).



\section{Automation Framework Elements}

This section describes the elements of the automation framework, with a focus on settings in the context of BO and Active Learning.



\subsection{Function $f(\boldsymbol{x})$}
\label{sec:function}

The function $y = f(\boldsymbol{x})$ is the function for which an optimal value should be found. In a power system setting, the function can be decomposed into a power system model and a post-processor. The power system model consists of a description of the electrical circuits and its controls, and takes as inputs (i) fixed parameters describing the model and (ii) parameters $\boldsymbol{x}$ which are to be optimized. The power system model should be solved using an appropriate method, e.g., an EMT-type solver. The solver delivers the value of the voltages and currents, and any other monitored parameters. For instance, an EMT-type solver delivers the voltages and currents as a function of time. The post-processor takes the outputs of the power system model and applies a function that maps those outputs to a desired output variable $f(\boldsymbol{x})$ or $y$. As an example, the output time-series can be translated into a single value by applying the $\max$-operator.

\subsection{Probabilistic surrogate model}
\label{sec:surrogate_model}

The probabilistic surrogate model $f_{\boldsymbol{\theta}}(\boldsymbol{x})$ is an important part of the method, and has to perform two important roles. First, it must provide an accurate fit of the function $f(\boldsymbol{x})$, especially in the region(s) close to $f(\boldsymbol{x^*})$. This ensures that the optimal value of $f(\boldsymbol{x})$ can be found via $f_{\boldsymbol{\theta}}(\boldsymbol{x})$. Second, it must provide a degree of certainty on the value of $f(\boldsymbol{x})$. This has a dual purpose: \emph{during} the optimization, the degree of uncertainty provides the automation algorithm with a guidance on trading off exploitation versus exploration.  \emph{After} the optimization, it provides the user a degree of certainty on the approximation of $f(\boldsymbol{x})$ by $f_{\boldsymbol{\theta}}(\boldsymbol{x})$. 

The selection of the probabilistic model is considered to be of primary importance to the functioning of the method~\cite{shahriari_taking_2016}. The probabilistic surrogate model should be selected such that it can approximate $f(\boldsymbol{x})$ well and it has a low computational complexity. For BO, there exists a wide variety of regression models, parametric and non-parametric. The selection between those could be guided by prior knowledge on the shape of $f(\boldsymbol{x})$. In case of non-parametric models, Gaussian Processes are a popular choice, with attractive properties with respect to flexibility in fitting and computational complexity. Nevertheless, in its most basic implementation, Gaussian Processes come with assumptions (all function values follow a jointly Gaussian distribution with the inputs, assumption of stationarity) and require careful selection of a kernel. We refer the reader to the literature for combinations or extensions on Gaussian Processes to alleviate these assumptions or for selecting kernels~\cite{duvenaud_automatic_2014}.

\subsection{Acquisition Function}
\label{sec:acquisition_function}

The second important choice for the automation framework is the choice of acquisition function. The acquisition function has the role of choosing the next sample $\boldsymbol{x}_{n+1}$ based on the previously evaluated samples ${(\boldsymbol{x}_1, y_1), (\boldsymbol{x}_2, y_2), ..., (\boldsymbol{x}_n, y_n)}$, in effect maximizing a certain utility function. The choice of the previous sample is guided by the probabilistic surrogate model, which indicates the regions where $f_{\boldsymbol{\theta}}(\boldsymbol{x})$ indicates high (or low) values of $f(\boldsymbol{x})$ on the one hand or indicates high uncertainty on $f(\boldsymbol{x})$. Choosing the right acquisition function involves trading off its computational complexity with its accuracy (and hence, efficiency in actually minimizing the samples needed).

Depending on the acquisition function, a different task is fulfilled. In the field of BO, several acquisition functions have been listed in~\cite{shahriari_taking_2016}. These acquisition functions balance exploration and exploitation to find the optimal value $y^*$ with a minimum of function evaluations. For example, if the objective is to find the optimal value of a function, the acquisition function can be based on expected improvement (EI), Gaussian Process upper confidence bound or predictive entropy search.

\section{Implementation}


The BO method used in this paper assumes the algorithmic form as outlined in Algorithm 1, taken from~\cite{shahriari_taking_2016}.  The function $f(\boldsymbol{x})$ is approximated by a Gaussian Process $GP$, which can use an exponentially squared kernel or a kernel of the Mat\'ern family, with the hyperparameters contained in $\boldsymbol{\theta}$. The surrogate model is denoted by $f_{\boldsymbol{\theta}}(\boldsymbol{x})$. For the optimization task, the acquisition function $\alpha(\boldsymbol{x};\mathcal{D}_n)$ is the expected improvement, whereas for the active learning task, the acquisition function is based on eq. (31) of~\cite{bect_sequential_2012}. To evaluate the acquisition function, the parameters $\boldsymbol{\theta}$ are either obtained using maximum likelihood, maximum a posteriori, or in a Bayesian Monte Carlo averaging method, marginalizing out the hyperparameters using the posterior distribution given the data. The objective function below uses the latter assumption:
\begin{equation}
  \alpha(\boldsymbol{x};\mathcal{D}_n) = \mathbb{E}_{\boldsymbol{\theta}|\mathcal{D}_n}\left[\mathbb{E}_{GP}\left[\max\{0, y^* - f_{\boldsymbol{\theta}}(\boldsymbol{x}) \} \right]\right],
  \label{eq:expected_improvement}
\end{equation}
in which $y^*$ is the current estimate of the optimal value.

\begin{algorithm}
\caption{Bayesian Optimization}\label{alg:cap}
\begin{algorithmic}
        \For{$n = 1, 2, ...$}
        \State select new $\boldsymbol{x}_{n+1}$ by optimizing $\alpha$
        \State $\boldsymbol{x}_{n+1} = \arg\max_{\boldsymbol{x}}\alpha(\boldsymbol{x};\mathcal{D}_n)$
        \State obtain $y_{n+1} = f(\boldsymbol{x})$
        \State augment $\mathcal{D}_{n+1} = \{\mathcal{D}_n, (\boldsymbol{x}_{n+1}, y_{n+1})\}$
        \State update statistical model
        \EndFor
\end{algorithmic}
\end{algorithm}

In the maximum likelihood estimation of $\boldsymbol{\theta}$, a maximum likelihood estimate $\boldsymbol{\hat{\theta}}$ is found by using the current data $\mathcal{D}_n$ to solve $\boldsymbol{\hat{\theta}} = \arg\max_{\boldsymbol{\theta}}\log\left(p(\boldsymbol{y}|\boldsymbol{x},\theta)\right)$. This can be one through plugging in the analytical expression (found in~\cite{shahriari_taking_2016}) for the log likelihood in a gradient descent optimizer. The inner expectation, denoted with subscript $GP$, can be written as $\alpha(\boldsymbol{x};\boldsymbol{\theta})$ and is then calculated analytically for $\boldsymbol{\hat{\theta}}$. This corresponds to the approach taken in~\cite{tanaka_application_2018}.

When a prior on the input parameters $\boldsymbol{\theta}$ is specified, the maximum a posteriori estimate may also be used. The maximum a posteriori estimate may be found by $\boldsymbol{\hat{\theta}} = \arg\max_{\boldsymbol{\theta}}\left(p(\boldsymbol{\theta})p(\boldsymbol{y}|\boldsymbol{x},\theta)\right)$~ \cite{garnett_bayesoptbook_2023}.

In the Bayesian Monte Carlo or averaging approach the two expectation operators in~\eqref{eq:expected_improvement} must be evaluated. To do so, a Markov Chain Monte Carlo method is used, which calculates the outer expected value by generating samples $\boldsymbol{\theta}^{(i)}$ that are distributed according to the posterior distribution $p(\boldsymbol{\theta}|\mathcal{D}_n)$:
\begin{equation}
\mathbb{E}_{\boldsymbol{\theta}|\mathcal{D}_n}\left[\alpha(\boldsymbol{x};\boldsymbol{\theta})\right] \approx \frac{1}{M}\displaystyle\sum_{i=1}^M\alpha(\boldsymbol{x};\boldsymbol{\theta}^{(i)}).
  \label{eq:mcmc}
\end{equation}



\section{Case Study}
\label{sec:studies}

The automation framework is applied to the assessment of transformer stresses during energization in the vicinity of underground cables. The application study borrows the parameters from~\cite{leterme_use_2021} in which the case study is outlined.

\subsection{Transformer Energization}

This case studies the overvoltages experienced during energization when resonances are matching harmonic frequencies, e.g., when energizing transformers in the vicinity of high-voltage cables. The transformer energization problem is extensively discussed in~\cite{cigre_wg_c4307_transformer_2014}. 

The objective of the study is to limit the probability on harmful resonant overvoltages during transformer energization to an acceptable level, e.g., less than five percent in~\cite{cigre_wg_c4307_transformer_2014}. Resonant overvoltages during tranformer energization may occur when the impedance of the connected grid has a high magnitude at a low-order harmonic frequency. Such a condition may be fulfilled if the grid impedance has a resonance peak at one of these frequencies. The actual occurrence of resonant overvoltages depends on the outcome of a stochastic process, where the epistemic input variable is the grid impedance and the aleatoric input variables are, amongst others, the switching time $t_{\text{s}}$ and transformer residual flux before energization, characterized by amplitude $\lambda_0$ and angle $\theta$.

In terms of computations, the transformer energization problem is challenging because of two main reasons: the computational effort for a single EMT simulation and the large number of simulations that must be performed. The simulation time of the EMT-type model is high even when reducing a grid to a frequency-dependent network equivalent; a case study presented in~\cite{vernay_application_2013} showed a simulation time of 4.7 seconds for a total simulation duration of 200 ms. In transformer energization studies, the simulation duration is typically in the range of seconds. The number of simulations required to achieve accurate results of a failure probability is in the order of thousands of simulations when using the Monte Carlo method. 

In this paper, we propose two solutions for this problem; first, finding the maximum voltage that may occur during transformer energization and second, classifying the transformer energization overvoltages according to a voltage threshold $U_{\text{thr}}$. The solution to the first problem helps in deciding whether a further probabilistic analysis is needed. The solution to the second problem can be used to actually calculate the probability of failure.

The case study makes use of a transformer connected to a reduced network as shown in~\cite{leterme_use_2021}. The transformer is modeled using the classical modeling approach with saturation modeled across the magnetizing inducance. The grid is modeled by a simple RLC-network, as also proposed by a system operator to mimic a transformer energized through a long cable \cite{Deschanvres2013TransientSP}.  The transformer parameters for the case study have been taken from~\cite{leterme_use_2021}. The RLC parameters are selected as $R= 1.32~\Omega$, $L= 50$~mH, $C= 50.6$~$\mu$F, corresponding to a resonance peak in the input impedance of amplitude 750 $\Omega$ at 100 Hz. As output of the simulation, the phase-to-phase voltage between phases a and b is chosen. The simulation model is implemented in EMT-type software, and executed with a time-step of 50 $\mu$s for a total simulation duration of 2.5 s.

The case study makes use of the input distributions for the aleatoric variables as proposed in~\cite{cigre_wg_c4307_transformer_2014}. Based on the sensitivity analysis performed in~\cite{martinez-duro_parameter_2012}, only the switching time, with a distribution of $t_s$ between 0 and 20~ms, and transformer residual flux, with a distribution of $\lambda_0$ between 0 and 0.8~pu and $\theta$ between 0 and 2$\pi$, are retained for further investigations. 
\begin{figure}[t]
  \centering
  \includegraphics{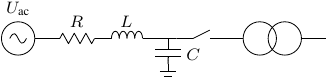}
  \caption{Electrical circuit for transformer energization analysis.}
  \label{fig:eleccircuit2}
\end{figure}



\subsubsection{Brute Force Benchmark Solution}

\begin{figure}[t]
  \centering
  \includegraphics[width=0.5\textwidth]{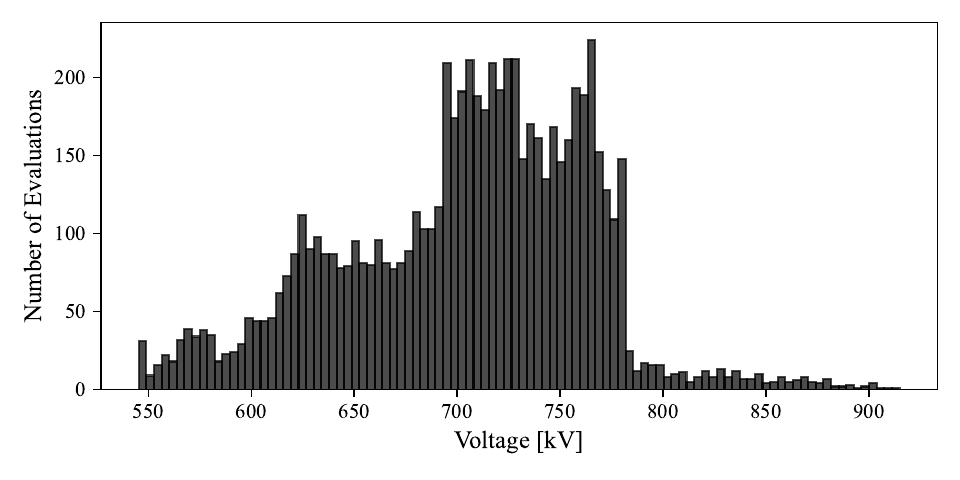}
  \caption{Brute Force Benchmark Solution}
  \label{fig:brute_force_benchmark}
\end{figure}

The distribution of the line-to-line voltages is skewed and has a maximum at $U_{\text{ab}}=916$ kV (Fig.~\ref{fig:brute_force_benchmark}). This distribution was obtained through a Monte Carlo Simulation that has drawn 7000 samples from the output distribution by sampling the input variables, running the EMT simulation and calculating the maximum voltage.

\subsubsection{Maximum Overvoltage (One Dimensional Input)}

In this case, only the switching time of the breakers is taken as stochastic input, taking a uniform distribution between 0 and 10 ms. The residual or remanent flux in the transformer (before switching) is set to 0. The optimization problem is therefore:
\begin{equation}
\begin{aligned}
\text{maximize} &~~~& {f(t_{\text{switch}})}\\
\text{subject to} &~~~& {0\leq t_{\text{switch}} \leq 10 \text{ms}},\\
\end{aligned}
\end{equation}
where $f(.)$ represents the function that applies the $\max$-operator to the output of the EMT-simulation. It should be noted that the switching time is bound to 10 ms instead of 20 ms because of the periodicity of the result due to the $\max$-operator.

To implement the BO, the Trieste toolbox~\cite{picheny_trieste_2023} is used. The regression model is the standard one offered within the toolbox, i.e., a Gaussian Process with the Mat\'ern 5/2-kernel. The inputs for the Gaussian Process are scaled to the unit box. The final parameters for the Gaussian Process are each time found using the Maximum A Posteriori (MAP) estimate.

\begin{figure}[t]
  \centering
  \includegraphics[width=0.5\textwidth]{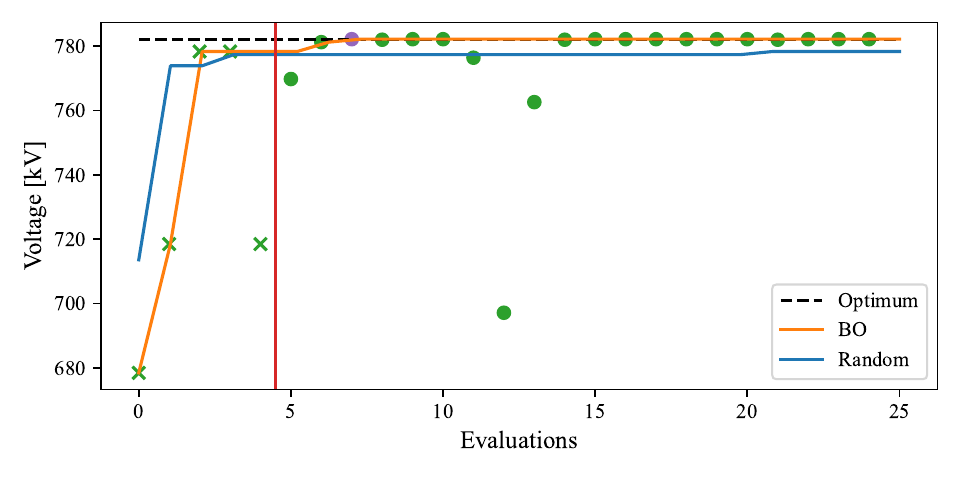}
  \caption{Optimal value against function evaluations (1D Input)}
  \label{fig:opt_value_iterations}
\end{figure}

\begin{figure}[t]
  \centering
  \includegraphics[width=0.5\textwidth]{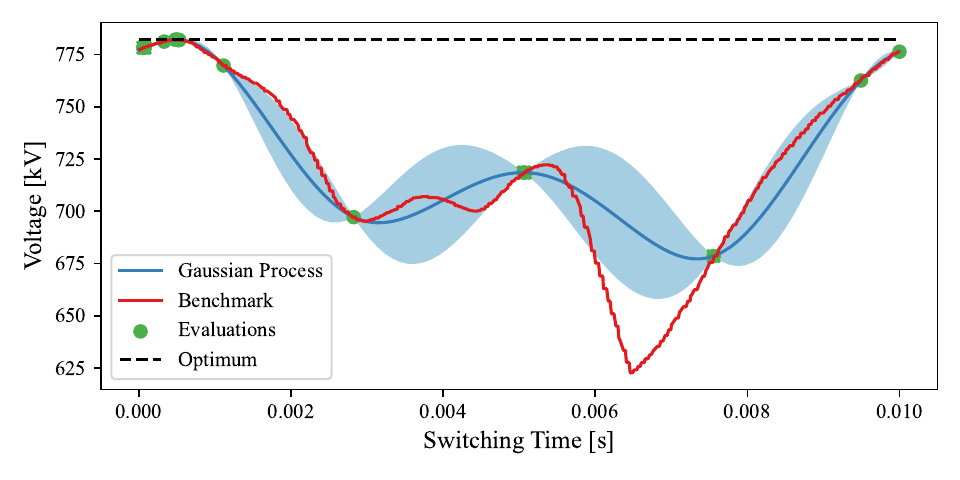}
  \caption{Surrogate model and benchmark for BO}
  \label{fig:opt_value_2}
\end{figure}

The BO converges to a value close to the maximum overvoltage within only a few iteration steps (Fig.~\ref{fig:opt_value_iterations}). In comparison, a random search did not reach the optimal value within 25 iterations. The flexibility of Gaussian Processes offers an advantage for the automation framework. Contrary to other fitting methods, e.g., polynomial regression or polynomial chaos, there is no need to sample the whole input space (Fig.~\ref{fig:opt_value_2}). Indeed, as shown in Fig.~\ref{fig:opt_value_2}, the Expected Improvement-criterion steers the sampling towards the areas close to the maxima (close to t=0 and t=10 ms), and does not aim to improve the model for other values (e.g., the interval t=[6,8] ms). Hence, the learned function $f_{\boldsymbol{\theta}}(\boldsymbol{x})$ matches the actual function $f(\boldsymbol{x})$ more closely for those values close to the optimum.

\subsubsection{Maximum Overvoltage (Three-Dimensional Input)}
\label{sec:three-d-case}

In this case, the stochastic inputs are not only the switching time, but also remanent flux amplitude $\lambda_0$ and angle $\theta$:
\begin{equation}
\begin{aligned}
\text{maximize} &~~~& {f(t_{\text{switch}}, \lambda_0, \theta)}\\
  \text{subject to} &~~~& 0\leq t_{\text{switch}} \leq 20 \text{ms},\\
                    &~~~& 0\leq \lambda_0 \leq 0.8 ,\\
                    &~~~& 0\leq \theta \leq 2\pi,\\
\end{aligned}
\end{equation}

\begin{figure}[t]
  \centering
  \includegraphics[width=0.5\textwidth]{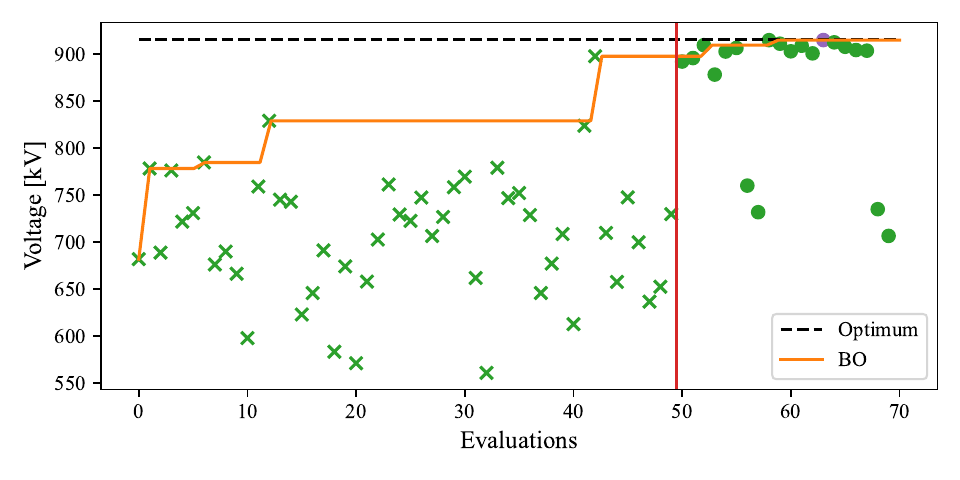}
  \caption{Optimal value against function evaluations (3D Input)}
  \label{fig:opt_value_iterations_3d}
\end{figure}

The Bayesian Optimizer is able to find the global optimum for the three-dimensional problem with a total of 70 function evaluations (Fig.~\ref{fig:opt_value_iterations_3d}). Compared with the brute force benchmark solution, the BO reduces the number of EMT simulations by a factor of 100. The ability of the optimizer to find the global optimum depends on the number of samples used to initialize the model. In this case, the optimizer has been initialized with 50 samples, indicated with a cross, while using 20 iterations for the actual optimization, indicated with a dot. If initialized with an insufficient number of samples, the optimizer may output a local optimum. It should be noted that the optimum differs from the previous case due to the additional aleatoric input variables.

Furthermore, the BO has been compared with alternative optimization methods which are all derivative-free and, in contrast to traditional gradient-based optimizers, do not require an exact mathematical description of the objective function \cite{doi:10.1137/1.9780898718768                                                                             }. The three optimization methods considered in this paper are (i) the Nelder-Mead (NM) algorithm, a deterministic local optimizer, (ii) the Differential Evolution (DE) algorithm, a non-deterministic global optimizer, and (iii) the Dual Annealing (DA) algorithm, a non-deterministic optimizer, which combines the global Simulated Annealing algorithm with a local search strategy. For detailed information on the algorithms, the reader is referred to \cite{Nelder.1965}, \cite{article}, \cite{TSALLIS1996395}, \cite{SZU1987157}. The alternative approaches have been implemented using the standard functions of the SciPy package. The maximum number of function evaluations, for all optimization methods, has been set to 100, consisting of both initialization samples and samples selected by the optimizer. To account for the dependency of the optimizers' outputs on the starting point, especially relevant for the local optimizer (NM), 10 optimization attempts with randomized starting points are conducted for each algorithm. Furthermore, the number of initialization samples for the BO has been varied, to analyze if a further reduction of the computational effort needed for initialization can be achieved.

\begin{figure}[t]
  \centering
  \includegraphics[width=0.5\textwidth]{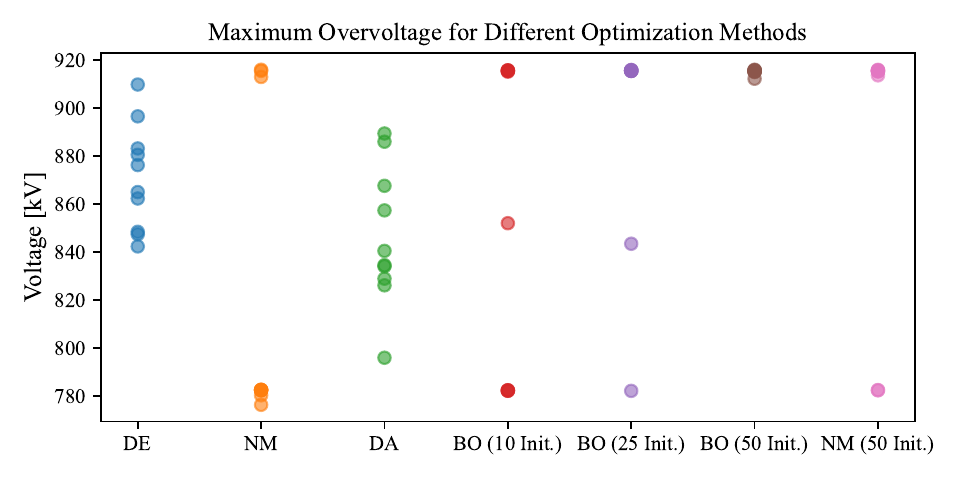}
  \caption{Maximum overvoltage for different optimization methods, showing the output for 10 randomly initiated attempts with a simulation budget of 100 function evalutions}
  \label{fig:opt_value_comparison_3d}
\end{figure}

Comparison of the optimizer's output after 100 function evaluations reveals that the BO, if initialized carefully, is able to outperform the other optimization algorithms (Fig.~\ref{fig:opt_value_comparison_3d}). For all attempts, the global optimizers DA and DE do not reach the optimal value within 100 function evaluations. The NM algorithm reaches the global optimum only for initialization close to that optimum and shows the largest spread in outputs. By initializing the NM algorithm with the optimum out of 50 randomly selected samples while using the remaining function evaluations for optimization, the optimizer's performance can be enhanced. In this case, the global optimum is reached in eight out of ten attempts. The performance of the BO varies with the number of initialization samples. For an initialization with ten samples, the BO reaches the global optimum in only five of the ten optimization attempts. By increasing the number of initial samples to 25, the global optimum is reached for 80 $\%$ of the attempts. With further increase to 50 samples, the BO obtains the global optmimum for all investigated optimization attempts. 


\begin{figure}[t]
  \centering
  \includegraphics[width=0.5\textwidth]{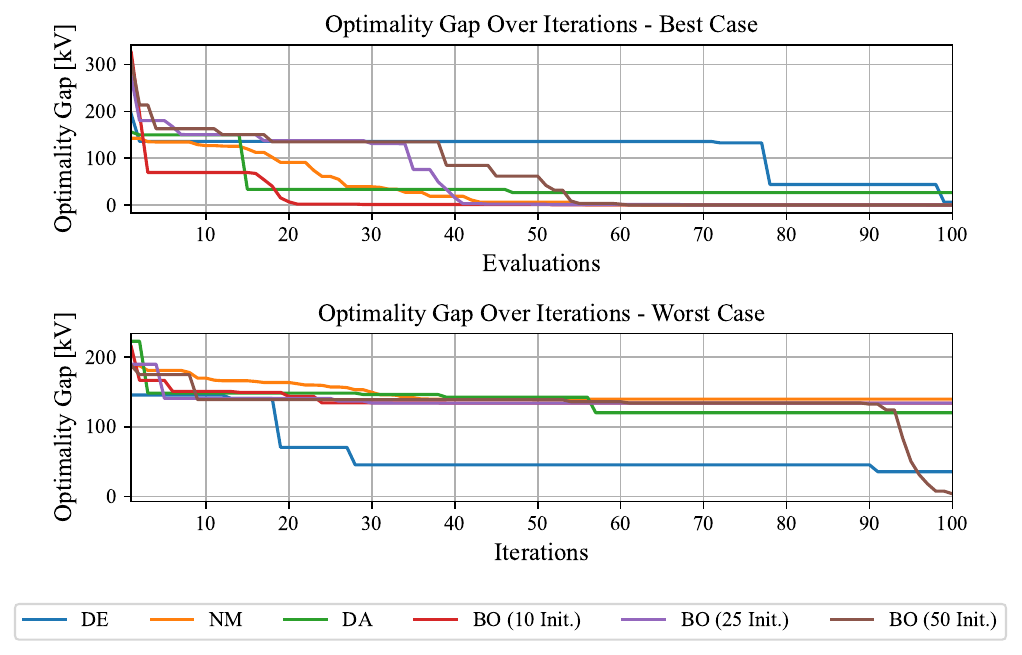}
  \caption{Optimality gap against number of function evaluations for different optimization methods for best and worst case of each of the 10 optimization attempts}
  \label{fig:optimality_gap_3d}
\end{figure}

Comparing the best case attempts for each algorithm, the BO, depending on the number of initialization samples, converges within a number of evaluations comparable to or lower than the NM algorithm (Fig.~\ref{fig:optimality_gap_3d}). For the worst case optimization attempts of each algorithm, only the Bayesian Optmization, if initialized with 50 samples, reaches a value close to the global maximum within the specified number of function evaluations.  



\subsubsection{Binary Classification}

In the transformer energization problem, unlike the maximum voltage, it may be of interest to find the probability of exceeding a voltage threshold. To solve this problem, the ``Bichon'' criterion is used, that is, putting $\delta$ to 1 in the acquisition function \eqref{eq:acq_fcn} proposed in \cite{bect_sequential_2012}. The parameter $\alpha$ is also set to 1.

\begin{equation}
\begin{aligned}
J \coloneq \mathbb{E}[\max(0, (\alpha s(x))^\delta - |T - m(x)|^\delta)]
  \label{eq:acq_fcn}
\end{aligned}
\end{equation}

For demonstration purposes and due to the simpler visualisation, the binary classification case is applied to the 1D-problem where only switching time is considered as a random variable in the first step. To start the classification, five samples are taken, and complemented with 10 more samples during the classification. The threshold voltage is set to 750 kV. Also for this case, the Trieste toolbox is used~\cite{picheny_trieste_2023}.

\begin{figure}[t]
  \centering
  \includegraphics[width=0.5\textwidth]{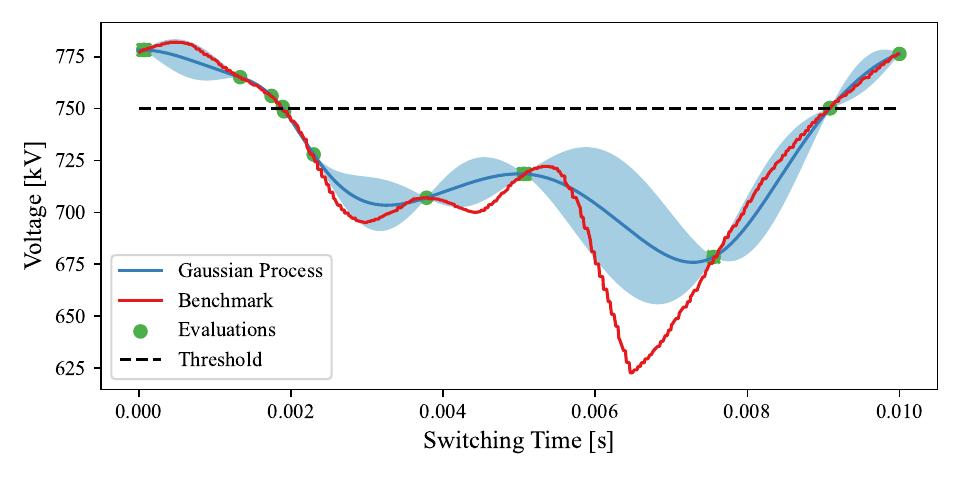}
  \caption{Surrogate model and benchmark for binary classification task}
  \label{fig:sumo_bench_binclass}
\end{figure}

With a relatively low number of samples, a relatively high accuracy in binary classification can be obtained. The binary classification using the 15 samples indicates a probability of samples above 750 kV of 29.5 $\%$, whereas the benchmark using a Monte Carlo approach with 7000 samples indicates 28.1 $\%$. It can be seen that the optimization routine concentrates the sampling points to those mainly around the threshold (Fig.~\ref{fig:sumo_bench_binclass}). Therefore, the surrogate model $f_{\boldsymbol{\theta}}(\boldsymbol{x})$ matches the actual function $f(\boldsymbol{x})$ more closely in areas around the specified threshold.  The light blue area refers to the 95 $\%$ confidence interval of the surrogate model $f_{\boldsymbol{\theta}}(\boldsymbol{x})$, which is the smallest in the areas close to the voltage threshold. It should be noted that while the actual function $f(\boldsymbol{x})$ is the same as in the 1D-optimization problem, the learned function $f_{\boldsymbol{\theta}}(\boldsymbol{x})$ differs from Fig.~\ref{fig:opt_value_2} due to the choice of acquisition function.


Finally, binary classification is applied to the three-dimensional problem. Based on the results from the 3D-optimization problem, the binary classification is initialized with 50 samples, complemented with 50 additional samples for the classification itself. The results in Table \ref{tab:3d_classification} show that with a limited number of 100 function evaluations the automation framework achieves binary classification with a relatively high accuracy for different voltage thresholds. By increasing the number of function evaluations to 200, the accuracy of the classification can be further enhanced.


\begin{table}
\centering
\caption{Binary classification for three-dimensional input}
\label{tab:3d_classification}
\begin{tblr}{
  row{odd} = {c},
  row{4} = {c},
  cell{1}{1} = {r=2}{},
  cell{1}{2} = {c=3}{},
  cell{2}{2} = {c},
  cell{2}{3} = {c},
  hline{1,3,6} = {-}{},
}
Threshold & Threshold Exceeding Probability &                      &                      \\
          & Benchmark                       & Bayesian (100 Eval.) & Bayesian (200 Eval.) \\
750 $kV$  & 23.69 $\%$                      & 20.81 $\%$           & 24.25 $\%$           \\
800 $kV$  & 2.743 $\%$                      & 2.457 $\%$           & 2.643 $\%$           \\
850 $kV$  & 0.943 $\%$                      & 0.829 $\%$           & 0.857 $\%$           
\end{tblr}
\end{table}

\section{Conclusion}

The automation framework proposed in this paper has the potential to reduce computational effort for problems involving computationally expensive power system simulations, while being suitable for a variety of tasks. For the transformer energization problem, maximum overvoltages or the probability of an overvoltage exceeding a threshold can be found in an efficient manner within 100 function evaluations, reducing the number of simulations compared to state-of-the-art approaches in industry by a factor of 10 to 100. Unlike other optimization routines, the Bayesian Optimization can be used for other tasks than optimization in a straightforward manner. This has been shown in the context of binary classification, where a change in the acquisition function enabled finding multiple intersection points between the non-linear function and a threshold.


The automation framework is mainly based on Bayesian Optimization, which requires choice of a surrogate model, acquisition function and number of evaluations. The acquisition functions used in this paper provided the desired outcome within a reasonable time. For this paper, standard surrogate models in the form of Gaussian Processes with Mat\'ern and exponentially squared kernels were used. It was found that the number of samples to initialize these models, especially with a larger number of input dimensions, was crucial for the correct functioning of the optimization. Further investigations on model selection and initialization are subject to future work.



\label{sec:conclusion}
\bibliographystyle{IEEEtran}
\bibliography{IEEEabrv,references}

\vspace*{5mm}

\noindent\textbf{M. Kuhn} received the B.Sc. and M.Sc. degree in electrical engineering, information technology, and computer engineering with a specialisation in electrical power engineering from RWTH Aachen University, Aachen, Germany, in 2021 and 2023, respectively. He is currently pursuing the Ph.D. degree in electrical engineering at RWTH Aachen University. His research interests include protection and insulation coordination of multi-terminal HVDC systems and their modeling for electromagnetic transient studies.\\

\noindent\textbf{E. Heylen} received the Ph.D. degree in electrical engineering on the topic
of power system reliability from Katholieke Universiteit Leuven (KU Leuven),
Leuven, Belgium. She is currently with Mindbrew BV, Berlaar, Belgium. She is a Guest Lecturer in the field of advanced topics on
control with Imperial College London, London, U.K. She was a Postdoctoral
Researcher with Imperial College London on the topic of data-driven techniques
for energy systems, in collaboration with National Grid ESO. Her research
interests include sustainable, secure, inclusive and affordable energy systems,
and how artificial intelligence can help to achieve this.\\

\noindent\textbf{W. Leterme} received the Ph.D. degree in electrical energy engineering from KU Leuven, Leuven, Belgium, in 2016.
He is currently a Professor of High Voltage Technology with the Institute
for High-Voltage Equipment and Grids, Digitalization and Energy Economics (IAEW) with RWTH Aachen, Aachen,
Germany. Previously, he was a Senior Researcher with KU Leuven, EnergyVille, Genk, Belgium. He was a Visiting Researcher with The University of
Manchester, Manchester, U.K., and Imperial College London, London, U.K., in 2015 and
2018, respectively. Dr. Leterme received the Ph.D. Fellowship from the Research Foundation Flanders. He received the Robert Sinave Thesis Prize by
KVBE/SRBE in 2021. His current research interests include high-voltage components and protection aspects of future power-electronic dominated ac/dc
systems.\\

\end{document}